\documentclass[twocolumn,showpacs,preprintnumbers,aps,prx,amsmath,amssymb,floatfix,superscriptaddress,longbibliography]{revtex4-1}

\usepackage{graphicx}
\usepackage{dcolumn}
\usepackage{bm}
\usepackage{color}
\usepackage{natbib}
\usepackage{textcomp}
\usepackage{url}           
\usepackage{appendix}
\usepackage{hyperref}
\hypersetup{colorlinks=true,urlcolor=blue,breaklinks=true,citecolor=blue,linkcolor=blue,pdfstartview=FitH,pdfpagemode=UseNone}

\newcommand{\ket}[1]{\left|#1\right\rangle}
\newcommand{\bra}[1]{\left\langle#1\right|}

\renewcommand{\eqref}[1]{Eq.~(\ref{#1})}

\def\NaRb{$^{23}$Na$^{87}$Rb }

\begin{document}

\title{Dipolar Collisions of Ultracold Ground-state Bosonic Molecules}

\author{Mingyang Guo} \thanks{Present address: 5. Physikalisches Institut and Center for Integrated Quantum Science and Technology, Universit\"at Stuttgart, Pfaffenwaldring 57, 70569 Stuttgart, Germany.}
\affiliation{Department of Physics, The Chinese University of Hong Kong, Hong Kong, China}
\author{Xin Ye}
\affiliation{Department of Physics, The Chinese University of Hong Kong, Hong Kong, China}
\author{Junyu He}
\affiliation{Department of Physics, The Chinese University of Hong Kong, Hong Kong, China}
\author{Maykel L. Gonz{\'a}lez-Mart{\'i}nez}
\affiliation{Laboratoire Aim\'{e} Cotton, CNRS, Universit\'{e} Paris-Sud, ENS Paris-Saclay, Universit\'{e} Paris-Saclay, 91405 Orsay Cedex, France}
\author{Romain Vexiau}
\affiliation{Laboratoire Aim\'{e} Cotton, CNRS, Universit\'{e} Paris-Sud, ENS Paris-Saclay, Universit\'{e} Paris-Saclay, 91405 Orsay Cedex, France}
\author{Goulven Qu\'{e}m\'{e}ner}
\affiliation{Laboratoire Aim\'{e} Cotton, CNRS, Universit\'{e} Paris-Sud, ENS Paris-Saclay, Universit\'{e} Paris-Saclay, 91405 Orsay Cedex, France}
\author{Dajun Wang}
\email{djwang@cuhk.edu.hk}
\affiliation{Department of Physics, The Chinese University of Hong Kong, Hong Kong, China}
\affiliation{The Chinese University of Hong Kong Shenzhen Research Institute, Shenzhen, China}

\date{\today}

\begin{abstract}

The dipolar collision between ultracold polar molecules is an important topic both by its own right from the fundamental point of view and for the successful exploration of many-body physics with strong and long-range dipolar interactions. Here, we report the investigation of collisions between ultracold ground-state sodium-rubidium molecules in electric fields with induced electric dipole moments as large as 0.7~D. We observe a step-wise enhancement of losses due to the coupling between different partial waves induced by the increasingly stronger anisotropic dipolar interactions. Varying the temperature of our sample, we find good agreement with theoretical loss rates assuming complex formation as the main loss process. Our results shed new light on the understanding of complex molecular collisions in the presence of strong dipolar interactions and also demonstrate the versatility of modifying molecular interactions with electric fields.

\end{abstract}

\maketitle

\section{Introduction}

Recent years have witnessed many breakthroughs in quantum simulation of strongly correlated many-body problems with dilute gases of ultracold particles~\cite{Bloch2008manybody,gross2017quantum}. While the majority of the current investigations are still carried out with atoms interacting via the short-range contact potential, there is a great interest in extending them to ultracold particles with long-range dipolar interactions. Popular systems for realizing dipolar interactions include special atomic species with large magnetic moments which can interact via magnetic dipole-dipole interactions~\cite{Griesmaier2005,Lu2011,Aikawa2012}, highly excited Rydberg atoms with gigantic electric dipole moments which can interact via strong electric dipole-dipole interactions~\cite{lukin2001,saffman2010,low2012}, and ultracold polar molecules (UPMs) which have permanent electric dipole moments (pEDMs) $\mu_0$ typically on the order of one atomic unit~\cite{Carr2009,bohn2017cold}.

While the dipole-dipole interaction between polar molecules is not as strong as that between Rydberg atoms, it is still much stronger than that between magnetic atoms. On the other hand, the lifetime of ground-state polar molecules is much longer than that of Rydberg atoms. Thus, UPMs hold great promises of broad applications in quantum simulation and quantum information~\cite{carr2009cold,trefzger2011ultracold,baranov2012condensed} and great experimental efforts have been devoted to bringing them into the quantum degenerate regime~\cite{ni2008high,takekoshi2014ultracold,molony2014creation,park2015ultracold,guo2016creation}. However, collisions between UPMs have proven to be much more complicated than those between ultracold atoms. In 2010, investigations carried out with fermionic $^{40}$K$^{87}$Rb molecules already observed rapid loss of molecules due to ultracold chemical reactions~\cite{ospelkaus2010quantum}. Very recently, by controlling the chemical reactivity of bosonic $^{23}$Na$^{87}$Rb molecules via vibrational excitation, a direct comparison has shown that the loss is similar with and without the two-body chemical reactions~\cite{ye2017chemical}.

The presence of the pEDM, the most important asset of UPMs, is also one of the main causes of the complicated collisions between them. This is already the case without the external electric fields when the induced electric dipole moment $\mu$ in the laboratory frame is zero, since the pEDM determines dominantly the long-range dispersion coefficient $C_6$ via the purely rotational transition contribution which is proportional to $\mu_0^4/B_v$ (with $B_v$ the rotational constant)~\cite{zuchowski2013van,lepers2013long,vexiau2015}. A substantial pEDM $\mu_0$ will lead to a rather large $C_6$. Together with the heavy mass of typical alkali dimers, this results in a very dense four-atom density of states (DOS) near the two-body collision threshold which favors a near unity probability of the two-molecule complex formation manifesting as loss of molecules, even in the absence of chemical reaction channels~\cite{mayle2013scattering,ye2017chemical}.

In electric fields, a non-zero $\mu$ in the laboratory frame allows two UPMs to interact via the direct dipole-dipole interaction $\mu^2(1-3\mathrm{cos}^2\theta)/(4\pi \epsilon_0 R^3)$, with $\theta$ the relative orientation between the molecules and $R$ the inter-molecular distance. The anisotropic nature of the dipole-dipole interaction can induce couplings between the relative motional angular momenta (partial waves $L$) and bring significant modifications to the long-range interaction potential and thus the collisions. Such dipolar effects between collisions of UPMs were first and so far only observed with relatively small $\mu$ in $^{40}$K$^{87}$Rb molecules which are identical fermions and chemically reactive~\cite{ni2010dipolar}.

In this work, we report our investigation of dipolar collisions in an ultracold sample of bosonic, chemically stable~\cite{zuchowski2010reactions} ground-state \NaRb molecules with $\mu$ up to 0.7~D. For the largest $\mu$, the characteristic length of the dipolar interaction $l_D=m \mu^2/4\pi\varepsilon_0\hbar^2$ reaches 0.75~$\mu$m, which is already comparable to the inter-molecular distance of 1.2~$\mu$m in our highest density samples. The characteristic energy scale $\varepsilon_D = \mu^2/4\pi\varepsilon_0 l_D^3\propto 1/m \mu^4$~\cite{Bohn2009} decreases to 7 nK, which is much smaller than our sample temperature $T$. Here, $m$ is the molecular mass and $\hbar$ is Planck constant divided by $2\pi$. Thus with $\mu$ increasing from 0 to 0.7~D, the collisions enter gradually the strongly dipolar regime in which the collisions are significantly different from the universal Wigner threshold regime, even though $T$ is only 100's nK. Experimentally, following the increase of the dipolar interaction, we observed this transition as pronounced enhancement of molecular losses with step-wise features as the result of the increasingly stronger modification to the effective interaction potentials of higher and higher partial waves.

\section{Experiment}

\subsection{Ground-state molecule preparation and polarization}

The optically trapped \NaRb samples were created by transferring weakly-bound Feshbach molecules~\cite{wang2015formation} to the absolute ground state (the $\ket{v = 0, J = 0, M_F = 3}$ level of the $X^1\Sigma^+$ electronic state, with $v$, $J$ and $M_F$ the vibrational, rotational, and total hyperfine quantum numbers)  via the stimulated Raman adiabatic passage~\cite{guo2016creation}. The Feshbach molecules were created by magnetoassociation of Na and Rb atoms both in their $\ket{F=1,m_F=1}$ hyperfine Zeeman level with the Feshbach resonance at 347.64~G\cite{wang2013observation,wang2015formation}. Here $F$ and $m_F$ are the quantum numbers of the atomic hyperfine level and its projection, respectively. For the population transfer, the $\ket{v=55,J=1}$ level of the $A^1\Sigma^+/b^3\Pi$ mixed state was used as the intermediate level~\cite{guo2016creation,guo2017high}. With a typical transfer efficiency of 93\%, up to $1.5\times 10^4$ optically trapped ground-state molecules could be created with near 100\% quantum state purity~\cite{guo2016creation}. We note that our transfer efficiency is already among the highest for similar experiments~\cite{ni2008high,takekoshi2014ultracold,molony2014creation,park2015ultracold,guo2016creation} and we have also verified previously that the small amount of imperfection has negligible effects on the collisions~\cite{ye2017chemical}. The cigar shaped optical trap was formed by crossing two 1064.4 nm laser beams at an angle of 20$^\circ$. The measured trap oscillation frequencies of the ground-state molecules were $(\omega_{x},\omega_{y},\omega_{z})=2\pi\times$(217(3),208(3),38(2)) Hz. With the largest sample, the calculated initial peak densities could reach $6\times10^{11}$ cm$^{-3}$.

Since the pEDM is along the molecular axis, it has to be oriented in the laboratory frame to induce the direct dipole-dipole interaction. To this end, we applied a DC electric field along the vertical direction and parallel to the external magnetic field of 335.2~G with two parallel Indium-Tin-Oxide glass electrodes placed outside the glass vacuum chamber~\cite{guo2016creation}. To prevent the problem of residual electric field from charge accumulation~\cite{moses2017new}, we limited the electric field to below 1~kV/cm. Thanks to the large pEDM for \NaRb and the relatively small rotational constant $B_v$, $\mu$ up to 0.7~D can already be induced with this moderate electric field.

\subsection{Loss rate constant measurement}

\begin{figure}[b]
	\centering
	\includegraphics[width=0.45\textwidth]{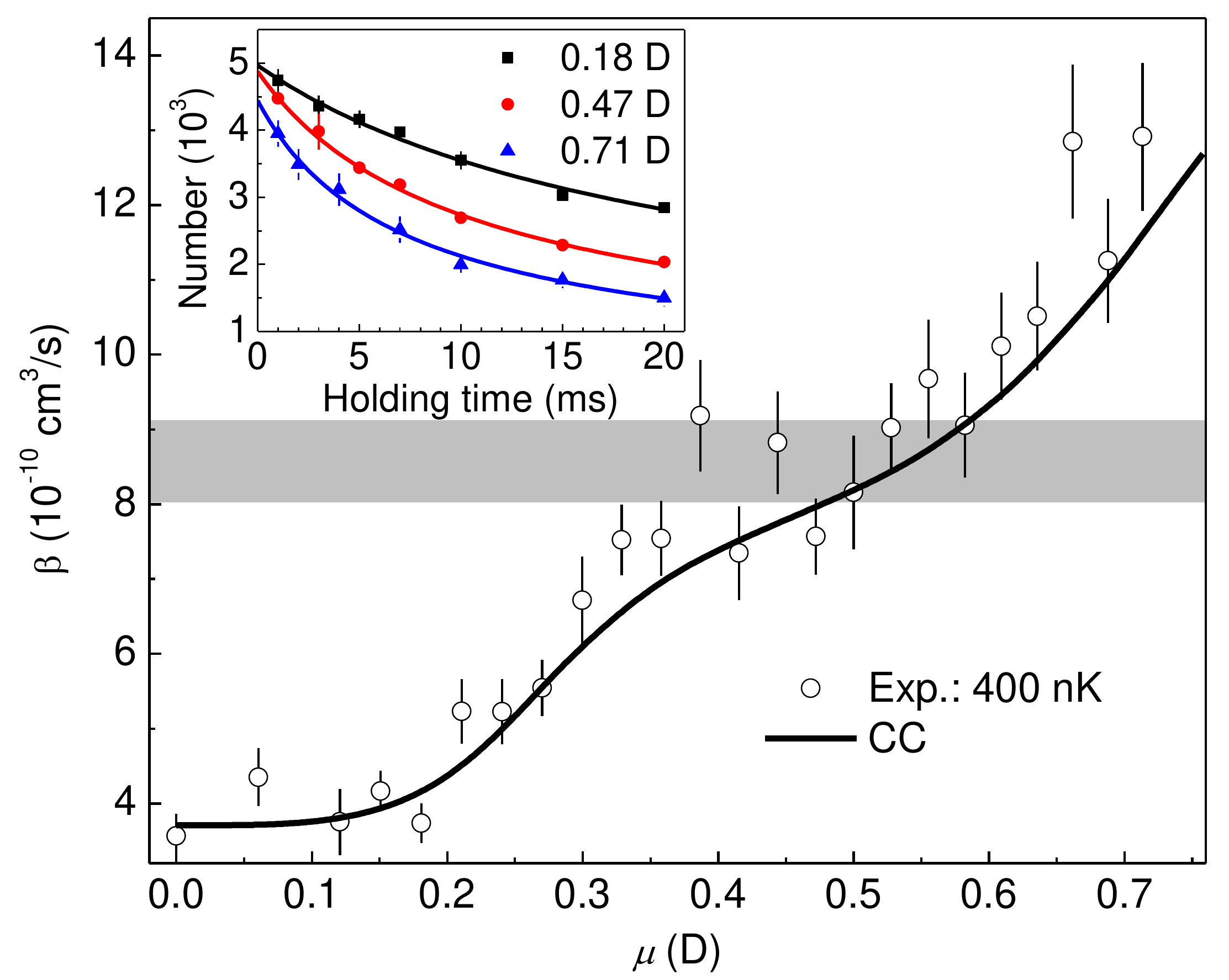}
	\caption{Two-body loss rate constant $\beta$ vs. the induced dipole moment $\mu$. The open circles are $\beta$ obtained from fitting the number loss and heating data simultaneously to Eq.~\ref{eq.1} with average sample temperatures of $T = 400$~nK. The error bars are from the fitting. The solid curve comes from a theoretical close-coupling calculation without adjustable parameters for the same sample temperature. The horizontal bar marks the inelastic $s$-wave unitarity limits of $T=400\pm50$~nK. Inset: example number loss measurements at different $\mu$ with error bars representing one standard deviation of typically three experimental runs.
	}
	\label{figure1}
	
\end{figure}

To investigate the dipolar effect in collisions, we measured the remaining number and the temperature of the ground-state molecule as a function of holding time for different $\mu$. As illustrated in the inset of Fig.~\ref{figure1}, following the increase of $\mu$, the loss of molecules accelerates substantially. Accompanying the number loss, the sample temperatures also increase rapidly. In our previous work with samples of the same temperature range but without the electric field~\cite{ye2017chemical}, we have observed that the collision is outside the Wigner regime and the loss rate constant $\beta$ is temperature dependent. For the current work, to avoid this complication in the data analysis, we limited the measurement time to less than 20 ms so that the temperature increase is less than 30\% thus $\beta$ can be treated approximately as a constant during the course of a single measurement.

We extracted $\beta$ for each $\mu$ by fitting the evolution of the molecule number $N$ and temperature $T$ simultaneously with the two-body loss model~\cite{Dalibard1998,ye2017chemical}
\begin{equation} \label{eq.1}
\begin{split}
\centering
	\frac{\mathrm{d}N}{\mathrm{d}t} & =-\beta A \frac{N^2}{T^{3/2}}\\
 	\frac{\mathrm{d}T}{\mathrm{d}t} & =\left(\frac{1}{4}+h_0\right)\beta A \frac{N}{T^{1/2}}.
\end{split}
\end{equation}
Here, $A=(m\bar{\omega}^2/4\pi k_B)^{3/2}$ is a constant, with $\bar{\omega} = (\omega_x \omega_y \omega_z)^{1/3}$ the mean trap frequency. In the temperature evolution, the 1/4 term is due to the potential energy. This can be understood from the fact that inelastic collision rates are faster near the center of the sample where the density is the highest but the potential energy is the lowest~\cite{Dalibard1998}. As a result, molecules removed from the trap have lower energies than the average. In addition, since the collision is outside the Wigner regime and $\beta$ depends on the collision energy~\cite{ye2017chemical}, following Ref.~\cite{Dalibard1998}, we also added the $h_0$ term to account for the kinetic energy contribution. For instance, the $s$-wave inelastic collision is faster at lower collision energies; thus, the collision prefers to remove molecules with smaller kinetic energies. We emphasize that in the Wigner regime, when $\beta$ is a constant, the $h_0$ term is not needed. As can be seen from the results in Fig.~\ref{figure1}, within the accessible $\mu$, $\beta$ increases in a step-wise manner. While $\beta$ changes little for $\mu$ from 0~D to $\approx$0.2~D, it rises rapidly as $\mu$ increases further. This is followed by a plateau from $\mu\approx 0.4$~D to 0.55~D. For $\mu > 0.55$~D, $\beta$ starts to increase again.

\section{Analysis}

\subsection{Qualitative understanding with the effective molecule-molecule interaction potentials}

To understand our observations, we look at the effective potential~\cite{quemener2011universalities,quemener2010strong}
\begin{equation}\label{eq3}
\begin{split}
	\bra{LM_L}V(R)\ket{L'M_L'}&=\left\{\frac{\hbar^2L(L+1)}{m R^2}-\frac{C_6}{R^6}\right\}\delta_{L,L'}\delta_{M_L,M_L'}\\
	&-\frac{C_3(L,L';M_L)}{R^3}\delta_{M_L,M_L'},
\end{split}
\end{equation}
in which the dipole-dipole interaction coefficient $C_3$ in Eq.~\ref{eq3} can be written in the partial wave basis as 
\begin{multline}
{C_3}(L,L';M_L)
= \frac{\mu^2}{4 \pi \varepsilon_0} 2 \ (-1)^{M_L} \ \sqrt{2L+1} \ \sqrt{2L'+1}   \\
\ \times\left( \begin{array}{ccc} L & 2 & L' \\ 0 & 0 & 0 \end{array} \right)
\ \left( \begin{array}{ccc} L & 2 & L' \\ -M_L & 0 & M_L \end{array} \right) \ .
\label{eq4}
\end{multline}
Here the brackets represent the $3j$-symbol which for identical bosons are non-zero only between the same $M_L$ components of even partial waves. This non-diagonal, dipolar interaction term in the effective potential mixes the same $M_L$ components of different even partial waves~\cite{quemener2011universalities,quemener2010strong}, e.g. between the $M_L = 0$ components of the $s$- ($L = 0$) and $d$-waves ($L = 2$), etc. In addition, there are also the diagonal, centrifugal term which vanishes for $s$-wave and the diagonal, isotropic van der Waals term characterized by the $C_6$ coefficient. Here, the $C_6$ coefficient includes a van der Waals term coming from the electronic structure, $C_6^{el} = 9018$~a.u.~\cite{vexiau2015} and another one coming from the rotational structure, $C_6^{rot} = \mu_0^4 / (192 \pi^3\epsilon_0^2 \hbar B_v) = 1.3184 \times 10^6$~a.u., calculated from the experimental values of $\mu_0 = 3.2(1)$~D~\cite{guo2016creation} and the rotational constant $B_v = 2.0896628(4)$~GHz~\cite{guo2017internal}. The total $C_6$, which is apparently dominated by the very large $C_6^{rot}$, amounts to $1.327 \times 10^6$ a.u. Since our molecules are identical bosons prepared in a single internal state, the scattering between them can only proceed via even partial waves.

\begin{figure}[bpt]
	\centering
	\includegraphics[width=0.9 \linewidth]{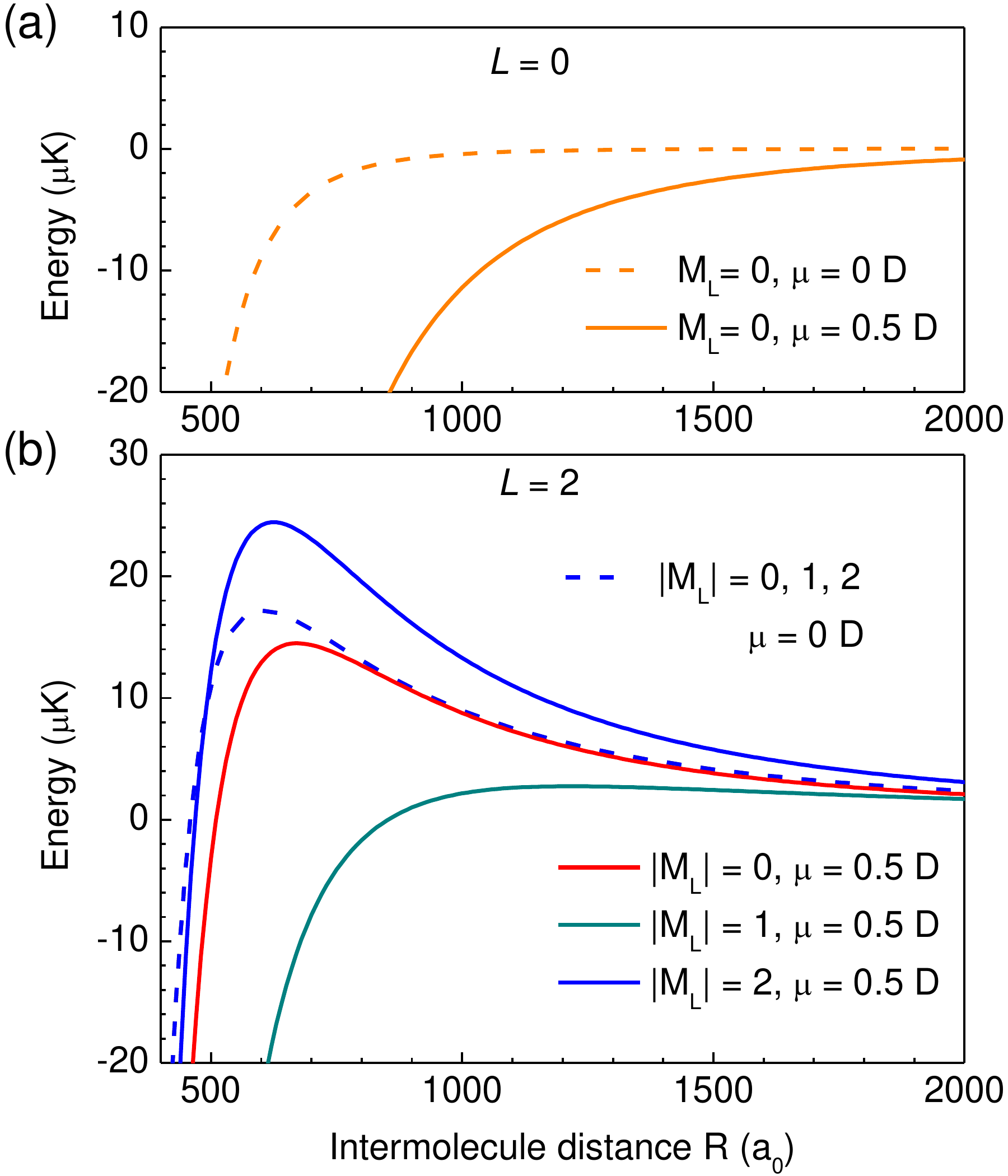}
	\caption{Dipolar modification to the inter-molecular interaction. 
	(a) The $s$-wave adiabatic potentials for $\mu=$0~D (dashed orange curve) and 0.5~D (solid orange curve). (b) The $d$-wave adiabatic potentials. For $\mu=$0~D (blue dashed curve), the different $M_L$ components overlap with each other. The $d$-wave barrier with a height of 17.2~$\mu$K is located at 600$a_0$, with $a_0$ the Bohr radius. For 0.5~D, the anisotropic dipolar interaction breaks the degeneracy which results in three different sets of potentials with $|M_L| = 0$ (solid red curve), 1 (solid cyan curve), and 2 (solid blue curve). The heights and the locations of the barriers are both modified substantially.}
	\label{figure2}
\end{figure}

At zero electric fields, the dipolar $C_3$ term vanishes and the effective potentials for $s$- and $d$-waves are depicted by the dashed curves in Fig.~\ref{figure2}(a) and (b), respectively. Due to the large total $C_6$, the barrierless $s$-wave effective potential has a large characteristic length $r_6=(m C_6/\hbar^2)^{1/4}$ of $720 a_0$ and a rather small characteristic energy $E_{6}=(\hbar^2/m r^2_6)$ of 3.0~$\mu$K~\cite{gao2010universal}, while the $d$-wave components ($|M_L| = 0,1,2$) coincide with each other with a barrier of 17.2~$\mu$K located at $R\approx 600 a_0$. Since our average sample temperature is $T\approx 400$~nK during the measurement, the scattering is dominated by $s$-wave.

\begin{figure*}[bpt]
	\centering
	\includegraphics[width=0.9 \linewidth]{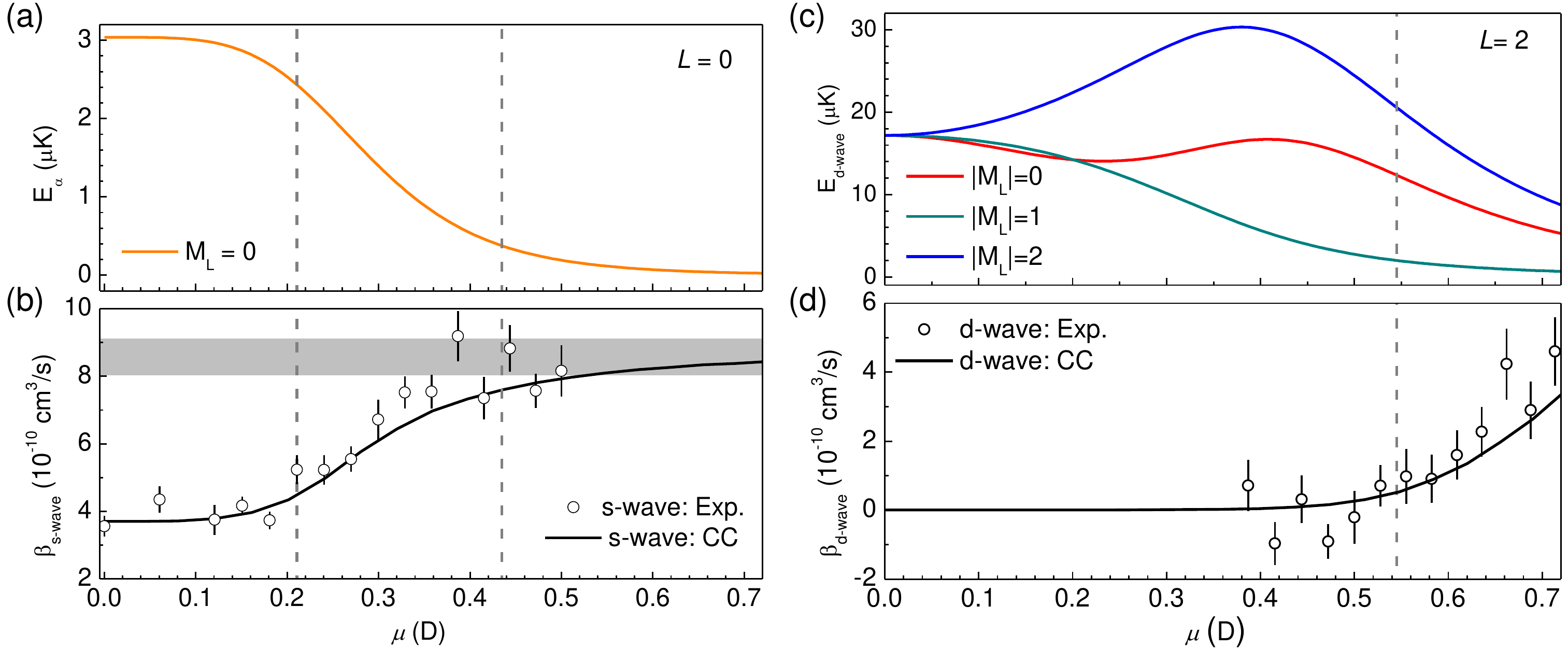}
	\caption{Dipolar collisions for individual partial waves. (a) and (b) are the generalized characteristic energy $E_{\alpha}$ and loss rate constant $\beta_{s-wave}$ vs. $\mu$ for $s$-wave. The vertical line near 0.22~D marks the position when the dipolar interaction starts to be comparable with the van der Waals interaction, while the vertical line near 0.44~D represents the point when $E_{\alpha}$ is lowered to $\approx T$. The gray horizontal bar in (b) is the inelastic $s$-wave unitarity limit for $T=400\pm 50$~nK. (c) illustrates the $d$-wave barrier heights $E_{d-wave}$ for different $|M_L|$ channels, and (d) shows $\beta_{d-wave}$ including contributions from all three $\vert M_L\vert$ channels with the $\vert M_L\vert = 1$ channel playing a dominant role. The losses from $\vert M_L\vert = 0$ and 2 are typically several hundred times slower than that from $\vert M_L\vert = 1$. The vertical bar near 0.55~D marks the position $E_{d-wave}\approx 5T$ for $|M_L|=1$.   
}
	\label{figure3}
\end{figure*}

In electric fields, when $\mu$ and thus the $C_3$ term in Eq.~\ref{eq3} are not zero, the effective potentials can be changed dramatically. As an example, Fig.~\ref{figure2} also shows the effective potentials for $\mu = 0.5$~D (solid curves) obtained by diagonalizing Eq.~\ref{eq3} for each $R$. Strictly speaking, with the coupling between different partial waves induced by the non-zero $C_3$ term, only $M_L$ remains a good quantum number but $L$ is not. However, for convenience, we will still distinguish the potentials with $L$ and keep the $s$-wave and $d$-wave notations. Comparing the two $s$-wave effective potentials in Fig.~\ref{figure2}(a), the $\mu = 0.5$~D potential gets more attractive due to the increasing strength of the dipolar interaction. In Fig.~\ref{figure2}(b), the $d$-wave components for $\mu=0.5$~D are split into three curves corresponding to $|M_L|=0, 1$, and $2$. Both the locations and the heights of the barriers for these three potentials are significantly modified from the $\mu =0$~D case. Especially, due to the coupling with the $g$-wave ($L = 4$), the potential barrier for the $\vert M_L \vert = 1$ component of the $d$-wave is vanishingly small which makes the $d$-wave scattering along this component favorable.

The modified $s$-wave potential can be represented by a generalized power law function $-C_\alpha/R^\alpha$ with the power index $\alpha$ varying from 6 to smaller values with increasing $\mu$~\cite{gao2008general}. Accordingly, the definitions of the characteristic length scale and energy can also be generalized to $r_{\alpha} = (m C_{\alpha}/\hbar^2)^{1/(\alpha-2)}$ and $E_{\alpha} = (\hbar^2/m r^2_{\alpha})$~\cite{gao2008general}. We have calculated the $s$-wave potentials for $\mu$ from 0 to 0.72~D. The $R = 400a_0$ to $2000a_0$ range of each potential is then modeled with the power law function with $\alpha$ and $C_\alpha$ as free parameters. Here, we chose a lower bound smaller than $r_6$ so that the van der Waals interaction can be fully taken into account. 

With $\alpha$ and $C_\alpha$ obtained from the fit, $r_{\alpha}$ and $E_{\alpha}$ can then be calculated for each $\mu$. The $E_{\alpha}$ vs. $\mu$ curve is shown in Fig.~\ref{figure3}(a). For a direct comparison, we replot in Fig.~\ref{figure3}(b) the measured loss rate constants for $\mu<0.5$~D, which we call $\beta_{s-wave}$ since they can be attributed mainly to $s$-wave scattering. In the van der Waals dominated region ($\mu < 0.2$~D), $E_{\alpha}$ is nearly a constant and $\beta_{s-wave}$ also changes little. Near 0.2~D, when the dipolar interaction becomes comparable to the van der Waals interaction and $E_{\alpha}$ starts to drop, $\beta_{s-wave}$ shows an apparent increase. For $\mu\approx0.44$~D, when $E_{\alpha}$ is lowered to be comparable to the sample temperature $T$, $\beta_{s-wave}$ reaches its maximum value bound by the unitarity limit $4\pi \hbar/m k$, with $k$ the scattering wave number determined by the sample temperature.

Similarly, as plotted in Fig.~\ref{figure3}(d), the $d$-wave contribution $\beta_{d-wave}$ for large $\mu$ can also be treated separately by subtracting out the unitarity limited $s$-wave rate constant. In this case, the characteristic energies are the barrier heights $E_{d-wave}$ of the $|M_L|$ channels which we extract from the calculated effective potential of $d$-wave for each $\mu$ (Fig.~\ref{figure3}(c)). Noticeably, the barrier height of the $|M_L|=1$ channel decreases with $\mu$ most rapidly and thus this channel contributes the most to the scattering. When the barrier height is still above the sample temperature, the loss rate is determined by the tunneling rate and thus small. However, when it finally lowered to a value comparable to the sample temperature at about 0.55~D, $\beta_{d-wave}$ starts to be significant and increases rapidly.

\subsection{Quantum close-coupled modeling}

Based on this understanding, we performed a quantum close-coupled (CC) calculation with $L$ up to 10. Following the highly-resonant scattering picture as a result of the very high four-atom DOS, at the short-range, a unit probability of collision complex formation is used in our model~\cite{mayle2013scattering,ye2017chemical}. Same as ultracold chemical reactions, the loss rate constant is then determined by the long-range interaction~\cite{ospelkaus2010quantum}. For zero electric field, we have verified previously~\cite{ye2017chemical} that the loss of non-reactive samples of \NaRb molecules is consistent with this model.

The CC calculation was performed using a time-independent quantum formalism, including the internal rotational structure of the NaRb molecule, an external electric field, and a partial wave expansion between them~\cite{Quemener2018}. Following the complex formation model, at short-range, a boundary condition is introduced so that when the two molecules meet, they are considered lost with a unit probability (see Ref.~\cite{Quemener2018} or~\cite{wang2015tuning} for more details). Asymptotic boundary conditions at long-range lead to the loss rate constant as a function of the applied electric field, and hence the corresponding induced dipole moment, for a given temperature. We imposed a short-range boundary condition at $R_{min} = 5a_0$ and a long-range boundary condition at $R_{max} = 10000a_0$. The loss rate constants are converged within 0.1\%.

Since our CC calculation includes the rotational structure, we input only the electronic coefficient $C_6^{el} = 9018$~a.u.~\cite{vexiau2015}, while the rotational coefficient $C_6^{rot}$ follows directly from the dipole-dipole interaction in the second order. We also used the aforementioned experimental $\mu_0$~\cite{guo2016creation} and $B_v$~\cite{guo2017high} values in the calculation. The $C_6^{rot}$ has a $\pm12\%$ uncertainty mainly from the uncertainty of the $\mu_0$ measurement. The uncertainty of $C_6^{el}$  can be safely ignored since it is more than 2 orders of magnitude smaller than $C_6^{rot}$~\cite{zuchowski2013van,lepers2013long,vexiau2015}.

The theoretical CC result without any adjustable parameters is shown as a solid black curve in Fig.~\ref{figure1} and agrees nicely with the measurement. Contributions from all $M_L$ components for $L$ up to 10 are included in order to obtain this curve. In addition, the calculated individual contributions from $s$-wave and $d$-wave, as shown by the solid black curves in Fig.~\ref{figure3}(b) and Fig.~\ref{figure3}(d), also agree well with the measured loss rate constants $\beta_{s-wave}$ and $\beta_{d-wave}$, respectively. We note that for the $\beta_{d-wave}$ calculation, $\vert M_L \vert = 0$, 1 and 2 channels are all included, although the contribution from $\vert M_L \vert = 1$ is typically several hundred times more than those from the other two channels.

\subsection{Temperature dependence}

\begin{figure}[b]
	\centering
	\includegraphics[width=0.9 \linewidth]{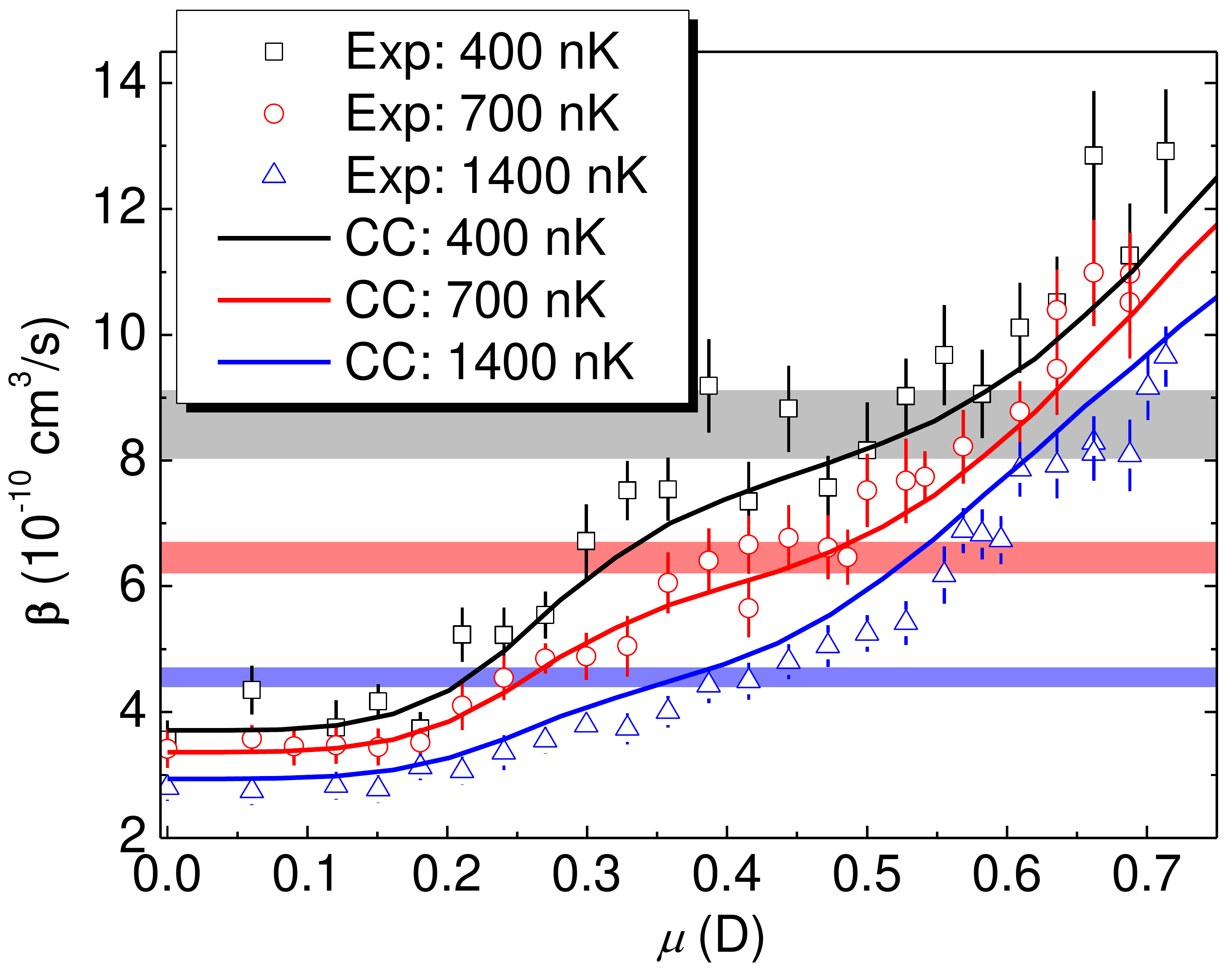}
	\caption{Dipolar collisions at different temperatures. The data points in black squares, red circles, and blue triangles are measured loss rates with sample temperatures of about 400~nK (same as in Fig.~\ref{figure1}), 700~nK and 1400~nK, respectively. The same color-coded solid curves are CC calculations with corresponding sample temperatures. The horizontal bars indicate the unitarity $s$-wave loss rates with temperature ranges of 400$\pm$50~nK, 700$\pm$50~nK and 1400$\pm$100~nK from top to bottom.
	}
	\label{figure4}
\end{figure}

For dipolar collisions in the Wigner regime, an approximately universal relation $\beta \propto \mu^{4(L+\frac{1}{2})}$~\cite{quemener2010strong,quemener2011universalities} has been verified experimentally for $p$-wave scattering in the $^{40}$K$^{87}$Rb system~\cite{ni2010dipolar}. This relation, however, is not expected to hold outside the Wigner regime. Indeed, power-law fittings to our data at $T=400$~nK give $\beta \propto \mu^{1.3(0.6)}$ for the $s$-wave collision (Fig.~\ref{figure3}(a)) and $\beta \propto \mu^{5.4(2.1)}$ for the $d$-wave collision (Fig.~\ref{figure3}(b)). Compared with the predicted relations $\beta \propto \mu^{2}$ and $\beta \propto \mu^{10}$, the $d$-wave collision with larger $\mu$ shows a much larger deviation. In addition, as can be seen in Fig.~\ref{figure4} which shows $\beta$ vs. $\mu$ for three different temperatures, $T=400, 700$, and $1400$~nK, the dependence of $\beta$ on $\mu$ changes obviously with $T$ and thus cannot be universal.

Several other important features can also be revealed from the measurements at different temperatures shown in Fig.~\ref{figure4}. Although near 0~D $\beta$ are different for different $T$, they all start to rise at nearly the same induced dipole moment of $\mu\simeq 0.2$~D since $E_{\alpha}$ is $T$ independent. However, for higher sample temperatures, the $s$-wave unitarity limit becomes lower and thus can be reached with smaller $\mu$. Meanwhile, with higher collision energy the $d$-wave barrier can also be overcome by the colliding molecules at smaller $\mu$. As a result, the boundary between the $s$-wave and the $d$-wave regimes becomes harder to distinguish. The CC calculations depicted by the three solid curves in Fig.~\ref{figure4} capture influences of the temperature very well.

\subsection{Discussion}
The good agreement between the CC calculation and the measurement indicates strongly that the short-range loss probability should be close to unity as we assumed in the model. This is also consistent with the lack of observable resonances for all three temperatures with the current steps of $\mu$. Both our present study with dipolar interactions and our previous one without dipolar interactions~\cite{ye2017chemical} are consistent with the model that complex formation with near unity probability is responsible for the loss of non-reactive ultracold \NaRb molecules. We further argue that this most probably is general for all other available non-reactive alkali polar molecules due to the similar four-atom DOS~\cite{mayle2013scattering}.

The above analysis comes with a caveat. Limited by the detection method, we cannot observe the complex directly. Furthermore, there must be some loss mechanism for the complex after its formation. Otherwise, it should either collide with another molecule or eventually dissociate back into two separated molecules~\cite{mayle2013scattering}. However, we have not observed any signatures of these possible post-complex-formation dynamics experimentally with or without the dipolar interactions~\cite{ye2017chemical}. Unfortunately, loss measurement in bulk samples is a rather cumbersome way to distinguish these possible contributions. It is thus very hard to claim whether it is these post-complex-formation dynamics or other unidentified mechanism are responsible for the loss of complexes. A much better method is to study molecular collisions in the single molecule level. Recently, the capability of forming single molecules in optical tweezers has been demonstrated~\cite{liu2018}.

\section{Conclusion}

In this work, we investigated the ultra-low energy dipolar collisions in a previously unexplored regime. We demonstrated how dipolar interactions modify profoundly the scattering between bosonic UPMs. While a strong dipolar interaction is often desirable for many applications, it also leads to collisions with significant contributions from multiple partial waves, typically $L=$ 0, 2, 4 ... for indistinguishable bosons and $L=$ 1, 3, 5 ... for indistinguishable fermions. This results in strongly anisotropic interactions between molecules. We note that for strong enough dipolar interactions, $e.g.$, as we have reached here, this kind of behavior should persist even in the zero-temperature limit. This is in stark contrast to the conventional ultracold collisions dictated by short-range van der Waals interactions which proceed via a unique single partial waves at low enough temperatures, typically $L=0$ for indistinguishable bosons and $L=1$ for indistinguishable fermions. This strongly dipolar regime achievable with UPMs is also beyond the scope of currently available dipolar quantum degenerate gases with magnetic dipolar interactions~\cite{Griesmaier2005,Lu2011,Aikawa2012} in which $\varepsilon_D$ is still much larger than the collision energy and thus universal threshold behavior can still be observed~\cite{Aikawa2014}.

For future investigation on many-body dipolar physics, suppression of the undesired loss is necessary. This may be achieved by dipolar interaction mediated long-range shielding with $J=1$ molecules in a specially chosen electric field~\cite{avdeenkov2006suppression,wang2015tuning,maykel2017} which for \NaRb is $\sim 4$~kV/cm. Although this field is out of reach for our current setup, it should be experimentally accessible with electrodes inside the vacuum chamber~\cite{moses2017new}. Alternatively, we can also prepare molecules in deep 3D optical lattice potentials to isolate them from each other~\cite{danzl2010ultracold,chotia2012long}. For a typical lattice constant of 532 nm, the currently achieved dipolar length $l_D$ is already enough to explore the strongly-correlated physics with significant non-local interaction effects.

\section{Acknowledgments}
We thank Bo Gao and Olivier Dulieu for insightful discussions. This work was supported by the COPOMOL project which is jointly funded by Hong Kong RGC (grant NO. A-CUHK403/13) and French Agence Nationale de la Recherche (grant NO. ANR-13-IS04-0004-01). The Hong Kong team was also supported by the RGC General Research Fund (grant NO. CUHK14301815) and the National Basic Research Program of China (grant NO. 2014CB921403).



%

\end{document}